\documentclass[prd,preprint,showpacs,showkeys,footinbib,floatfix]{revtex4}
\usepackage{graphicx} \usepackage{amssymb} \usepackage{amsmath} \baselineskip6pt
 \def\bea{\begin{eqnarray}}
\def\eea{\end{eqnarray}} 

\newcommand{\dipole}{\ensuremath{q \bar{q}}}
\newcommand{\pom}{\ensuremath{I\!\!P}}             
\begin{document}
\title{\hskip4in   \\ 
Hadronic Interactions of  Ultra-High Energy Photons with Protons and Light Nuclei 
in the Dipole Picture} \date{\today{}}
\author{T.C. Rogers}  \email[E-mail:]{rogers@phys.psu.edu} 
\author{M.I. Strikman} \email[E-mail:]{strikman@phys.psu.edu}
\affiliation{Department of Physics, Pennsylvania State University,\\
University Park, PA  16802, USA } 

\begin{abstract}
We apply the dipole formalism that has been developed to describe
low-x deep inelastic scattering
to the case of ultra-high energy real photons with nucleon and
nuclear targets.   
We hope that there will be future modeling applications in 
high-energy particle  astrophysics. 
We modify the dipole model of McDermott, Frankfurt, Guzey, and Strikman (MFGS) by fixing the
cross section at the maximum value allowed by the unitarity constraint
whenever the dipole model would otherwise predict a unitarity violation.  We observe
that, under reasonable assumptions, a significant fraction of the real
photon cross section results from dipole interactions where the QCD
coupling constant is small, and that the MFGS model is
consistent with the Froissart bound. 
The resulting model predicts a rise
of the cross section of about a factor of 12 when the the photon energy
is increased from $10^{3}$~GeV to $10^{12}$~GeV.
We extend the
analysis to the case of scattering off
a $^{12}$C target. We find that, due to the low thickness
of the light nuclei, unitarity for the scattering off individual nucleons plays a larger role than for the 
scattering off the nucleus as a whole. At the same time the proximity to the black disk limit results in a 
substantial increase of the amount of nuclear shadowing.  This, in turn, slows down the rate of increase of the total cross section with 
energy as compared to the proton case. As a result we  find that the $^{12}$C nuclear cross section rises by about a factor of 
7 when the photon energy is increased from $10^{3}$~GeV to $10^{12}$~GeV.
We also find that the fraction of the cross section due to production of charm reaches 30\% for the highest considered 
energies with a $^{12}$C target.  
\end{abstract}
\pacs{12.38-t, 13.85.Tp}
\keywords{QCD, Phenomenological Models}

\maketitle

\section{Introduction}
\label{sec:intro}     
There is currently an interest in the types of
showers induced by ultra-high energy (UHE) cosmic ray neutrinos~\cite{Klein:2004er}  
which will be relevant to  
the Anita, Auger and Icecube experiments.  
The resulting 
showers are presumed to be initiated by the Bremsstrahlung photons
radiated from the electron produced in the initial reaction,
\begin{equation}
\nu_{e} + A \longrightarrow e + X,
\end{equation}
where $A$ is a nucleus in the target medium, and $X$ is a 
hadronic jet produced in the initial reaction.
Due to the Landau-Pomeranchuk-Migdal (LPM)~\cite{lpm} effect, at UHE soft electromagnetic radiation is
suppressed and
most of the energy of the electron is transfered directly to the photon.
Furthermore, the cross section for $e^{+} e^{-}$ pair production drops
at UHE and the hadronic interaction between the photon and the 
target nuclei may dominate while the electromagnetic interaction
becomes negligible.  The suppression due to the LPM effect becomes stronger depending on the density 
of the target medium.  For a general overview of the LPM effect and electromagnetic suppression various media, see Ref.\cite{Klein:1998du}.
Furthermore, the shapes of showers may depend upon whether they are dominantly electromagnetic or hadronic~\cite{Acosta:1991cb}.  
Along these lines, it also important to determine what fraction of showers are due to charmed particles.  This is 
important for IceCube and MACRO because an increase in the number of charmed mesons in the initial reaction will increase the number of high energy
muons seen in experiments, but the contribution of high energy muons is one factor used to determine the composition of 
cosmic rays.  In addition, charmed particles contribute to the flux of atmospheric neutrinos.  Thus, experiments  
will need to take this contribution into account in searches for diffuse astrophysical neutrinos.  
For example, constraints on the UHE neutrino flux are sensitive (see, e.g., Ref.~\cite{Barwick:2005hn}) are sensitive to constraints on UHE photon cross sections.   

Another possible source of UHE cosmic photons is the decay of extremely
masses exotic particles, topological defects, and Z-burst models~\cite{topdown}. Particles with masses as high as 
$10^{26}$~eV may explain the observation of
super-GZK energy cosmic rays~\cite{Aloisio:2003xj}.  Indeed, the calculations in Ref.~\cite{Aloisio:2003xj}
have shown, using both standard QCD and supersymmetric QCD, that a large part of the
spectrum in the decay of super-massive particles consists of photons.
A characteristic of these ``top down'' models is the existence of a large photon flux in cosmic rays.  The ratio of 
protons to photons in the primary interaction for the production of showers can be used to distinguish between various top-down scenarios.
It is necessary, in order to address the issues above, to place upper limits on the growth of the real photon cross section.
Upper limits have been placed on the the fraction of 
primary photons at $26 \%$~\cite{Risse:2005hi} by analysis at the Auger observatory, but those upper limits 
are sensitive to the photon-nucleus interaction.
The main purpose of this paper is to combine the 
dipole picture 
(in the limit that the incident photon is real) with unitarity constraints and phenomenological
expectations to  
estimate 
the actual growth of the real photon
cross section with target nucleons/nuclei at UHE. 
We allow the cross section to grow with  energy as fast as
possible under the constraints of the S-channel unitarity, and  
thereby we 
at the very least place upper limits on the growth of photon-nucleus cross sections. 
It is   worth emphasizing here that one cannot simply use a smooth extrapolation
of the cross section to higher energies by assuming (as it is often done for the case
of hadron-hadron scattering) a parameterization of the cross section inspired
by the Froissart bound of the form, $\sigma^{\gamma N}_{tot}= a + b \ln^2(s/s_0)$.  
Asymptotically, the photon - hadron cross section can grow \emph{faster} than the rate of 
growth
supplied by the Froissart bound due to 
the the fact that the photon wave function is non-normalizable.  In fact, as we will discuss, the
rate of growth with energy of the photon-hadron cross section may be as fast as $\ln^{3}E_{\gamma}$ \cite{Frankfurt:2001nt,footnote2}.  
Because of the uncertainties involved in describing astronomically high energy photon behavior, the strategy that we take in this
paper is to provide meaningful upper limits on the growth of the cross section based on the 
observation 
 that the true hadronic cross section
of the photon will be tamed in any realistic theory relative to what is predicted in a low order perturbative QCD description.  Therefore, a low 
order perturbative QCD description that is only corrected for definite unitarity violations supplies a absolute upper limit.  We purposefully 
avoid including a more precise description of higher order affects (using, e.g., saturation, color glass condesnsate, etc...) in order to avoid
introducing any model dependent taming effects into the description of the high energy photon.  Therefore, we most likely overestimate the cross section 
by a significant fraction, but we will see that the upper limits obtained are nevertheless useful since they provide stronger constraints than more complex models of
the high energy behavior.

Dipole models in strong interaction physics 
describe the interaction of a high energy virtual photon 
with a hadronic target by representing the photon by a
distribution over hadronic Fock states of varying sizes in the light-cone
formalism.  This picture has become popular for studying deep inelastic 
scattering in the low-x limit.  
It is a
\emph{dipole} model because the very small size configurations are
usually modeled by a small quark/anti-quark pair, though larger size
Fock states are not necessarily dipole configurations and should be
thought of as more general hadronic states 
containing, for example, intrinsic gluon fields. The idea that hadrons should contain configurations interacting with different strength emerged 
first in the context of discussion of the inelastic diffraction ~\cite{FP,Good:1960ba,Miettinen:1978jb}. 
This idea reemerged in connection with the introduction of the two gluon exchange model for the strong interaction \cite{Low:1975sv,Nussinov:mw}.
In this model the interaction of a small hadronic system with a large hadronic system is obviously proportional to the transverse size squared of this system\cite{Low:1975sv}.  
In this context the impact parameter  representation of the scattering amplitude  first introduced for high energy processes by Cheng and Wu \cite{Cheng} in their studies of the high energy  
QED turned out to be useful.
The relationship between the smallness of the interaction with nucleons, the small size of the incident configuration (color transparency), and Bjorken scaling was emphasized in \cite{FS88} where it was 
demonstrated that small and large size configurations give comparable contributions to $F_{2}(x,Q^2)$ at small x and 
$Q^2\sim few\, GeV^2$ - the QCD aligned jet model.
The use of the eikonal model with two gluon exchange
 and the Cheng and Wu representation for the impact parameter photon wave function (extended to finite $Q^2$) was used to build a model of nuclear shadowing 
 \cite{Nikolaev:1990ja}.
 
It was pointed out in \cite{Blaettel:1993rd} that the cross section of the dipole nucleon interaction within the leading log approximation is actually proportional to the gluon density at high virtuality and small x, resulting in a fast growth of the total cross section.  The resulting dipole-nucleon cross sections were used  
to take into account finite $Q^2$ corrections to exclusive vector meson production in DIS  \cite{Frankfurt:1995jw}.
Splitting of the dipoles into systems of dipoles was studied within the BFKL approximation in \cite{Mueller:1994jq}.  The dipole model was applied to the description of HERA DIS data by
Golec-Biernat and Wusthoff~\cite{Golec-Biernat:1998js} who introduced a parameterization of the dipole cross section inspired by the eikonal model
which ensured that the total cross section of the interaction of dipoles of any size would reach a finite limiting value at high energies. 
 
A more realistic model was introduced 
by McDermott, Frankfurt, Guzey and Strikman (MFGS) in
Ref.~\cite{McDermott:1999fa}.  In  \cite{McDermott:1999fa} (the MFGS model) the
cross section for small dipole sizes was constrained to satisfy the perturbative QCD expression for the dipole - nucleon interaction, while for large sizes a  
growth consistent with the pattern of the pion - nucleon interaction  was imposed.
An important  advantage  of the model is that one can 
adjust the behavior of large size and small size configurations 
independently.  
The MFGS model was extended to an
impact  parameter analysis in Ref.~\cite{Rogers:2003vi}, and other
impact parameter analyses were done earlier in
Ref.~\cite{Munier:2001nr,Shoshi:2002in}.
 
At small Bjorken-x, the gluon
distribution dominates over the quark distribution in the nucleon
target, and the behavior of the small size dipole cross section is 
successfully predicted in
leading-twist perturbative QCD (pQCD).   However, at very small values
of Bjorken-x, the leading-twist gluon distribution becomes
unreasonably large and qualitatively new physics is expected to
dominate.  The usual assumption is that, whatever QCD mechanism is responsible 
for extremely low-x behavior,
scattering of a dipole of a given size at fixed impact parameter at sufficiently high energy occurs at or near the limit allowed by unitarity.    Applications of
the dipole model have thus far been used mainly in searches for these
qualitatively new regimes in QCD.  
Within the dipole picture,
at fixed large $Q^2$ and sufficiently small x
 the distribution of sizes
is sharply peaked around small size quark-antiquark pairs.  
For small photon virtualities,
$Q^{2}$, and for the energies $E_{\gamma} \le 10^3$~GeV
the distribution in sizes is  dominated mostly
by large sizes which lie far from the pQCD regime.  However, for extremely 
high energy photons (relevant to cosmic rays), pQCD predicts a very rapid rise in the basic
cross section with an increase in dipole size.  At asymptotically large
photon energies, finite size configurations rapidly reach the unitarity
bound and certainly lie far outside the applicability of ordinary
leading-twist pQCD.  Thus, we expect that for real 
photons with astronomically large
energies, a sizable contribution to the total $\gamma N$ cross section
may arise from small size dipoles whose amplitudes in impact parameter space are close to the the saturation limit for a 
wide range of impact parameters. 
To address quantitatively the issue of the real photon - nucleon  or the real photon - light nucleus interaction at cosmic ray energies we need to use a model which 
naturally matches with pQCD for small dipole sizes as it is the small size dipoles which will play the crucial role in the analysis of maximal possible growth of the cross section. Hence we will 
  use the MFGS model. The consistency of a particular dipole model can be
checked by considering both the real photon limit and the limit of
extremely high photon energies where the total cross section should
start to exhibit behavior consistent with S-channel unitarity. 
 We will
demonstrate this consistency within the MFGS model in the present 
paper.
  
The paper is organized as follows:
In Sec.~\ref{sec:section1} we discuss the modifications of the MFGS model necessary to 
apply it to real photon - nucleon scattering.
In Sec.~\ref{sec:section2} we evaluate the rate of 
growth of the $\gamma$-proton cross section at UHE and we indicate the advantages
and limitations of the current approach. We estimate, within the model, the fraction of the total cross section due to the 
diffractive processes and find it to be quite close to the black disk limit of 50\%. We also demonstrate that nearly 25\% of 
the total cross section at the highest energies comes from production of leading charm (the interaction of the photon in 
the $c\bar c$ component at photon energies of $10^{12}$~GeV).  
For a target $^{12}$C nucleus, this ratio rises to around 30\% due to differences in the degree of shadowing for charm and light quark dipoles.
In Sec.~\ref{sec:section3} we extend the
results of Sec.~\ref{sec:section2} to the case of a $^{12}$C target. We find that blindly extending the procedure we used for the 
proton target (imposing an upper limit of unity on the nuclear profile function) leads to the paradoxical result that the ratio of the
nuclear and nucleon cross sections exceed the total number of nucleons, A.  We explain that this is a consequence of the small 
thickness of the target.  
The main result of our investigations is the general observation that taming is necessary for the 
elementary cross sections of the hadronic subprocesses in $\gamma$-N scattering, and that, due to the large amount of diffractive scattering for 
the hadronic subprocesses in $\gamma$-N scattering, there 
is a significant amount of shadowing in $\gamma$-A scattering.
We develop and we summarize 
our results in the conclusion.  In App.~\ref{sec:smallx} we indicate the parameterization that
we used to extrapolate the gluon parton density to extremely small $x$, and in App.~\ref{sec:froissart} we
give a general description of the energy dependence of UHE cross sections based on the properties of the photon wave function and energy dependence of the dipole - nucleon interaction.

\section{The Photo-Production Limit}
\label{sec:section1}
In the perturbative QCD  dipole model, the total (virtual) photon - nucleon cross section due to 
the interaction of the small size configurations with a target is written as the
convolution product of a basic perturbative cross section for the interaction
of the 
hadronic Fock component of the photon  with the transverse and 
longitudinal light-cone wave
functions of the photon, $\psi_{L,T}(z,d)$, see e.g. Refs.~(\cite{Frankfurt:1995jw,footnote1}):
\begin{eqnarray}
\sigma^{\gamma N}_{T,L}(x,Q^{2}) = \int_{0}^{1} dz \int d^{2} {\bf d} \left| \psi_{T,L}(z,d)
\right|^{2} \hat{\sigma}_{tot}(d,x^{\prime})\,.
\label{eq:equation1}
\end{eqnarray}
If the 4-momentum of the incident photon is $q$, then, in the conventional notation, $Q^{2} \equiv -q^{2}$.
Of course, the longitudinal component of the photon wave function
vanishes in the limit that the photon is real, $Q^{2} \rightarrow 0$, and the transverse
component becomes a unique function of the quark momentum fraction,
$z$, and the hadronic size, $d$.  The transversely polarized photon's light-cone wave function
is,
\begin{equation}
\left| \psi_{T} (z,d) \right|^{2} = \frac{3}{2 \pi^{2}} \alpha_{e.m.} 
\sum^{n_{f}}_{q = 1} e_{q}^{2} [ (z^{2} + (1 - z)^{2}) \epsilon^{2} K_{1}^{2}(\epsilon d) 
+ m_{q}^{2} K_{0}^{2}(\epsilon d)], \label{eq:equation1b}
\end{equation}
where the sum is over $n_{f}$ active quark flavors, $m_{q}$ is the quark mass,
the $K$'s are the modified Bessel functions of the second kind, $e_{q}$ is the fractional 
charge of quark $q$, and $\epsilon^{2} = Q^{2} (z(1 -z)) + m_{q}^{2}$. 
(See Ref.~\cite{Rogers:2004rw} for discussion of 
appropriately dealing with quark masses.)  
For UHE photons, Eq.~(\ref{eq:equation1b}) will include a term
for the light quarks ($m_{q} \approx .3$~GeV) as well as a term for charm
($m_{q} \approx 1.5$~GeV).  More massive quarks are strongly suppressed by 
the light-cone wave function and are neglected in the present analysis (but see Sec.~\ref{sec:section2} for more discussion of
the heavier quarks).  
The energy dependence in Eq.~(\ref{eq:equation1}) enters through the Bjorken-x variable commonly used in deep inelastic 
scattering:
\begin{equation}
x \equiv \frac{Q^{2}}{2P \cdot q}, \nonumber
\end{equation}
where $P$ is the 4-momentum of the target nucleon.
The MFGS model specifies the 
small size and large size cross sections and a scheme for interpolating
between the two.
In the original formulation of the
MFGS model~\cite{McDermott:1999fa}, $x$ and $Q^{2}$ are taken as input
for the calculation of a particular cross section.  Note in Eq.~(\ref{eq:equation1}) that the
value of Bjorken-$x$, $x^{\prime}$, used in the basic dipole-nucleon cross section
is not the  same as the
external value of Bjorken-$x$.  
A relationship between $x$ and $x^{\prime}$ must be supplied by the specific model, and 
for the MFGS model $x^{\prime}$ is an effective average Bjorken-x for the dipole-nucleon 
scattering subprocess,   
\begin{equation}
x^{\prime}  =  \frac{Q^{2}}{2 P \cdot q} \left( 1  +
            \frac{4m_{q}^{2}}{Q^{2}} \right)  \left( 1 + \frac{0.75
            \lambda }{d^{2}( Q^{2}  + 4m_{q}^{2})} \right). \label{eq:equation2}
\end{equation}
We will now give a 
brief review of where this expression comes from. 
The method of
calculating $x^{\prime}$ used in Eq.~(\ref{eq:equation2}) is derived  
for the small size components of the
photon where the photon fluctuates into a $q \bar{q}$ pair which
then exchanges a single gluon with the target.  
It can be verified that the important
values of $x^{\prime}$ are centered around $x^{\prime} \approx 1.75 x$ in the usual deep inelastic Bjorken scaling regime. 
For a detailed derivation of the relationship between $x$ and $x^{\prime}$, see Ref.~\cite{McDermott:1999fa}.
By ignoring aligned jet configurations $(z \rightarrow 1,0)$, we can write the expression for the invariant mass of the produced $\dipole$ system 
as,
\begin{equation}
M_{\dipole}^{2} \approx 4 (m_{q}^{2} + {\bf k}_{\perp}^{2} ). \label{eq:equation2b}
\end{equation}
Using Eq.~(\ref{eq:equation2b}) along inside the exact expression for $x^{\prime}$,
\begin{equation}
x^{\prime} = \frac{Q^{2} + M_{\dipole}^{2}}{s + Q^{2}},
\end{equation}
we arrive at the expression,
\begin{equation}
\label{equation2c}
x^{\prime} = x \left( 1 + \frac{4 m_{q}^{2}}{Q^{2}} \right) \left(1 + \frac{{\bf k}_{\perp}^{2} }{(Q^{2} + 4 m_{q}^{2})} \right).
\end{equation}  
The typical numerical value ${\bf k}_{\perp}^{2}$ contributing to the integral  should be similar to the average 
$\langle {\bf k}_{\perp} ^{2}\rangle \sim \lambda /d^{2}$ that
we use in sampling the gluon distribution.  In order to reproduce the $x^{\prime} \sim 1.75 x$ behavior in the Bjorken limit, we use ${\bf k}_{\perp}^{2} = .75 \lambda/d^{2}$.
This reproduces Eq.~(\ref{eq:equation2})
The model is well-defined (finite) in
the photo-production limit, $Q^{2} \rightarrow 0$; but the
external value of $x$ vanishes for all energies when $Q^{2}=0$ and is
clearly not appropriate as input to the basic cross section.  Rather,
we would like to fix $Q^{2} = 0$ and specify $E_{\gamma}$.
To this end, we note how the original value of $x^{\prime}$ used in
Ref.~\cite{McDermott:1999fa} behaves for $Q^{2} \rightarrow 0$:
\begin{eqnarray}
x^{\prime}  \stackrel{Q^{2} \rightarrow 0}{=}  \frac{4m_{q}^{2}}{2 P \cdot q}
            \left(1 + \frac{0.75 \lambda}{4 d^{2} m_{q}^{2}} \right).
            \label{eq:equation3}
\end{eqnarray}
The constant, $\lambda$,  relates
the size of the hadronic configuration to the virtuality, $\bar{Q}^{2}$, of a particular quantum 
fluctuation of the
photon through the relation, $\bar{Q}^{2} = \lambda / d^{2}$.  Note that, even when  $x \rightarrow 0$, the
effective value, $x^{\prime}$, is large when $d \rightarrow 0$.  We emphasize that, although the appearance of Eq.~(\ref{eq:equation3}) seems 
at first glance to be somewhat ad hoc for use at UHE, it is sufficient for our present purposes to simply note that $x^{\prime}$ is directly proportional 
to $x$ in this limit and inversely proportional to $d^{2}$.  Estimates of the sensitivity to other factors in expression~(\ref{eq:equation3}) can be obtained 
by varying the quark mass and the parameter $\lambda$.  

In the original formulation of the MFGS model, the large size
configurations were characterized by growth with energy that mimicked
the $\ln^2(\frac{W^{2}}{W_{0}^{2}})$ growth of the pion-nucleon total
cross section with $W_{0}^{2} = 400$~GeV$^{2}$ (or $x_{0} \approx 0.01$ for DIS kinematics).
That is, it was assumed in Ref.~\cite{McDermott:1999fa} that for large sizes,
\begin{eqnarray}
\sigma(x^{\prime},Q^{2}) = \sigma_{\pi N} (x^{\prime},Q^{2}) = 23.78
\, \mathrm{mb}  \left( \frac{x_{0}}{x} \right)^{0.08}. \label{eq:equation4}
\end{eqnarray}
This behavior for the pion cross section was extracted from data in
Ref.~\cite{Burq:1981nv}.  Furthermore, soft Pomeron exchange leads to a factor of
$e^{\alpha^{\prime} t \frac{d^{2}}{d_{\pi}^{2}} \ln(\frac{x_{0}}{x})}$ with $\alpha^{\prime} = .25$~GeV$^{-2}$ and
$d_{\pi} = .65$~fm.
in the scattering amplitude.  In the case of the virtual
photon discussed in Refs.~\cite{McDermott:1999fa,Rogers:2003vi}, the
external value of $x$ was used for the large size behavior.  Clearly,
this is inappropriate in the photo-production limit where $x=0$.

In fact, we will now argue that the appropriate value of $x$ to use for the
large size configurations is the same $x^{\prime}$ (Eq.~(\ref{eq:equation2})) that was used for
the small size behavior in Ref.~\cite{McDermott:1999fa}.   To
see this, consider the large $W^{2}$, fixed $q^{2}$, limit of the pion-nucleon
scattering cross-section (considering, for the moment, $q$ to be the {\bf pion} 4-momentum):
\begin{eqnarray}
W^{2} & = & (q + P)^{2} \\ \nonumber  & = & M^{2} + 2P \cdot q + q^{2}
      \\ \nonumber  & \stackrel{x \rightarrow 0}{=} & 2P \cdot q.  \label{eq:equation5}
\end{eqnarray}
Then,
\begin{eqnarray}
\ln\left( \frac{W^{2}}{W_{0}^{2}} \right)  & = & \ln\left( \frac{2P
      \cdot q}{2P \cdot q_{0}} \right) \nonumber \\ & = & \ln\left(
      \frac{x_{0}}{x} \right)  + \ln\left( \frac{Q^{2}}{Q_{0}^{2}}
      \right).                         \label{eq:equation6}
\end{eqnarray}
Where $Q_{0}$ is defined so that $x_{0} \equiv Q_{0}^{2}/(2P \cdot
q_{0}) = .01$.  If we regard the pion mass as a particular value for the 
photon virtuality, then we see that Eq.~(\ref{eq:equation6}) generalizes to any small
photon virtuality $Q^{2}$ and small Bjorken scaling variable, $x$. 
On the other hand, in the limit of small $Q^{2}$ and
non-vanishing size, $d$, Eqs.~\ref{eq:equation2},~\ref{eq:equation3}
and~\ref{eq:equation5} yield,
\begin{eqnarray}
\ln\left( \frac{x_{0}^{\prime}}{x^{\prime}} \right) & = & \ln\left(
                                         \frac{2P \cdot q}{2P \cdot
                                         q_{0}} \right)\\
                                         \label{eq:equation7} & = &
                                         \ln\left(
                                         \frac{W^{2}}{W_{0}^{2}}
                                         \right),
                                         \label{eq:equation8}
\end{eqnarray}
where $x^{\prime}_{0}$ corresponds to taking $Q \rightarrow Q_{0}$.
In other words, to truly mimic the behavior of the pion-nucleon cross
section, one should use the effective Bjorken-x that was used in
sampling the gluon distribution.  Comparing Eqs.~\ref{eq:equation6}
and~\ref{eq:equation7}, we also have
\begin{eqnarray}
  \ln\left( \frac{x_{0}^{\prime}}{x^{\prime}} \right) - \ln\left(
\frac{x_{0}}{x} \right) = \ln\left( \frac{Q^{2}}{Q_{0}^{2}}
\right). \label{eq:equation9}
\end{eqnarray}
Thus, if one uses the external values of $x$ in
Eq.~(\ref{eq:equation4}) then
one over-estimates the large size cross section by a factor of
$(Q_{0}^{2}/Q^{2})^{.08}$ which diverges in the limit of a real photon. 
Finally, we can determine the value of $x^{\prime}_{0}$ by noting
that $W_{0}^{2} = 400$~GeV$^2$ implies $Q_{0}^{2} = 4$~GeV$^2$.
Therefore, from Eq.~(\ref{eq:equation3}), we have,
\begin{equation}
x^{\prime}_{0} = x_{0} (1 + m_{q}^{2}) \left( 1 + \frac{.75 \lambda}{4 d^{2} (1 + m_{q}^{2})} \right). \label{eq:xo}
\end{equation}
The second term, $m_{q}^{2}$, in parentheses in Ref.~\ref{eq:xo} is implicitly 
divided by $1$~GeV$^{2}$ so that it is unitless.
Notice that the difference between $x_{0}$ and $x_{0}^{\prime}$ is 
only significant for small $d$ or large $m_{q}$.  
The conclusion of this section is that it is appropriate to use  the
effective $x^{\prime}$ given in Eq.~(\ref{eq:equation2}) for all values
of  $Q^{2}$ and that Eq.~(\ref{eq:equation2}) \emph{must} be used in the
photo-production limit.  Furthermore, $x^{\prime}$ is calculated
unambiguously from Eq.~(\ref{eq:equation3}) and $E_{\gamma}$ and with
the condition that the target nucleon is at rest so that $P \cdot q = M E_{\gamma}$.  For the rest of this paper, we assume that the target
nucleon is at rest and we specify $E_{\gamma}$  as input for the dipole
model of the real photon.

\section{Growth of the Cross-Section at Very High Energies}
\label{sec:section2}

When one considers the profile function as in
Ref.~\cite{Rogers:2003vi}, or the S-matrix as is done in
Ref.~\cite{Munier:2001nr}, one finds that the unitarity constraint is
usually violated at very low values of Bjorken-$x$ (Ultra-High Energy).  This sort of
behavior is evidence of breaking of the leading twist approximation
and indicates the onset of qualitatively new
QCD phenomena. 
  However, the small size configurations of the UHE real photon
wildly violate the unitarity constraint due to the rapid growth of 
the perturbative expression,
\begin{equation}
\hat{\sigma}_{pQCD}(d,x^{\prime}) = \frac{\pi^{2}}{3} 
d^{2} \alpha_{s}(\bar{Q}^{2}) x^{\prime} g_{N}(x^{\prime},\bar{Q}^{2}), \label{eq:pertex}
\end{equation}
at small $x^{\prime}$
even when the increase of the radius of the interaction with energy is taken into account.  Therefore, it becomes necessary to 
introduce some new assumptions about the unitarity violating components
of the total cross section in order to make some sense of the UHE behavior.
We choose to adhere to the
usual assumption that at UHE, in the region where small
size perturbative methods begin to break down, the cross section is at
(or near) the limit set by unitarity.  This assumption is supported by our studies of the amplitudes of the dipole - nucleon interaction at HERA energies~\cite{Rogers:2003vi}.  
For reference, the hadron-proton amplitude that was used in Ref.~\cite{Rogers:2003vi} and which we use in this paper is given by,
\begin{equation}
F_{hN}(s,t) = i s \hat{\sigma}_{tot} \frac{1}{ (1  -
  t/M^{2}(d^{2}))^{2}} \frac{1}{1 - t d^{2} /d_{\pi}^{2} m_{2}^{2}}
  e^{ \alpha^{\prime} \frac{d^{2} t}{d_{\pi}^{2}}  \ln \frac{x_{0}}{x}}
\label{eq:ourmodel},
\end{equation}
with,
\begin{equation}
M^{2}(d^{2}) = \left\{ \begin{array}{ll} m_{1}^{2} - (m_{1}^{2} -
             m_{0}^{2}) \frac{d^{2}}{d_{\pi}^{2}}  &
             ,\mbox{\hspace{2mm}$d \leq d_{\pi}$} \\
                                                               
            m_{0}^{2} & ,\mbox{\hspace{2mm}otherwise}
                       \end{array} \right. \,. \label{eq:M}
\end{equation}
(See Ref.~\cite{Rogers:2003vi} for the detailed procedure for obtaining this.)   For these equations,
$m_{1}^{2} \approx 1.1$~GeV$^2$, $m_{0}^{2}  \approx 0.7$~GeV$^2$ and
$m_{2}^{2} \approx 0.6$~GeV$^2$.  The  typical size of the pion is
$d_{\pi} \approx 0.65$~fm,  $\alpha^{\prime}$ is $0.25$~GeV$^{-2}$, and $x_{0} = .01$.
We use Eq.~(\ref{eq:ourmodel}) to provide continuity with our previous publications while noting that the
details are not important for our current purposes; any reasonable model of the $t$-dependence that matches correctly 
to lower energy behavior is sufficient.  At UHE, the dominant effect is from the diffusion factors.  By making a large 
estimate for the amount of diffusion 
($\alpha^{\prime}$),
 we over estimate the amount of spreading in impact parameter space.  Within 
our approach to unitarity taming, as will
become clear, this over estimate of the diffusion is consistent with the goal of obtaining an upper limit to the growth of the cross section.

Very large size configurations
tend not to violate the unitarity constraint  in a gross way
- the total cross section
for large size configurations increases slowly with energy, but there
is suppression of the impact parameter  at small $b$ with increasing energy due to diffusion
consistent with the Donnachie-Landshoff soft Pomeron~\cite{Donnachie:en}.  
In fact, the fit of  ~\cite{Donnachie:en} does violate S-channel unitarity for
LHC energies and above. However, this leads to a very small effect for the total cross sections for the energies we discuss here.
We can enforce the unitarity assumption about the
high energy behavior if we re-calculate the 
total photo-production cross section by using the basic cross section
of the MFGS model with linear interpolation everywhere \emph{except} where
the unitarity limit is violated.  When the unitarity limit is
violated, we set the cross section equal to the maximum value allowed
by unitarity.  More specifically, we first decompose the the basic cross section 
in terms of the
impact parameter representation of the amplitude (usually called the profile function) using the optical theorem.  
Let $F_{h N}(E_{\gamma},l)$ be the hadron-nucleon amplitude and let
$l$ be the 4-momentum exchanged in the subprocess.
Then, we calculate, 
\begin{eqnarray}
\sigma^{\gamma N}_{tot}(E_{\gamma}) = 2 \int_{0}^{1} dz \int d^{2} {\bf{d}} \, \left|
\psi_{T}(z,d) \right|^{2} \int d^{2} {\bf{b}} \, \Gamma(E_{\gamma},b,d)\, ,
\label{eq:equation10}
\end{eqnarray}       
where $\Gamma(E_{\gamma},b,d)$ is the profile function for a configuration of size $d$.
The hadronic profile function that we use is given by the usual definition,
\begin{equation}
\Gamma(E_{\gamma},b,d) = \frac{1}{2 i s (2 \pi)^{2}} \int d^{2} {\mathbf l} \, e^{i {\mathbf b} \cdot {\mathbf l}} F_{h N}(E_{\gamma}, l), \label{eq:profile}
\end{equation}
so long as it is less than one.  If it exceeds unity, then we explicitly
reset $\Gamma(E_{\gamma},b,d) = 1.0$.  
Note that in the limit of $d \rightarrow 0$, both $x^{\prime}$ and the
photon wavefunction become mass independent.  
As we shall see, the distribution of hadronic sizes involved in the interaction becomes
more and more sharply peaked around small sizes as $E_{\gamma} \rightarrow \infty$.
Thus, in the UHE limit,
the sum in Eq.~(\ref{eq:equation1b}) will contain significant contributions from all massive strongly interacting 
particles.  The energies that we consider in this paper are still not high enough to include all 
heavy particles, but, as we will discuss, there is a significant charm contribution.
We include the charm contribution in the calculation of the cross section and note that the resulting
hadronic showers will likely consist of a significant number of charmed particles like D-mesons.
The contribution from the bottom quark will be suppressed relative to the charm due to 
its larger mass and smaller electromagnetic coupling.  We neglect the contribution from bottom and all
heavier particles.

\begin{figure}
\rotatebox{270}{\includegraphics[scale=0.60]{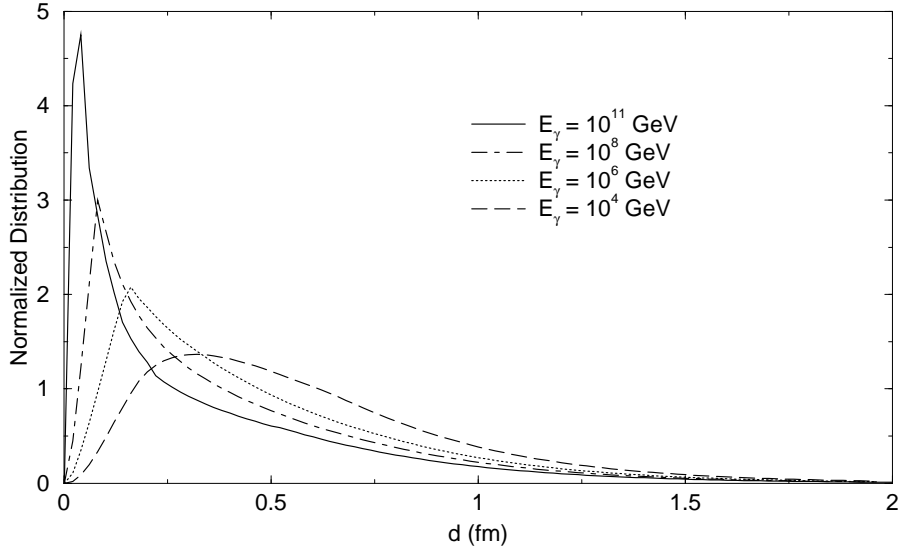}}
\caption{\label{fig:figure2}The distribution of the integrand in
Eq.~(\ref{eq:equation1}) over hadronic sizes.  Here, the unitarity constraint, $\Gamma \leq 1$, 
is explicitly enforced.  Nevertheless, the distribution becomes sharply peaked around small hadronic
sizes for UHE photons.}
\end{figure}
 
\begin{figure}
\rotatebox{270}{\includegraphics[scale=0.60]{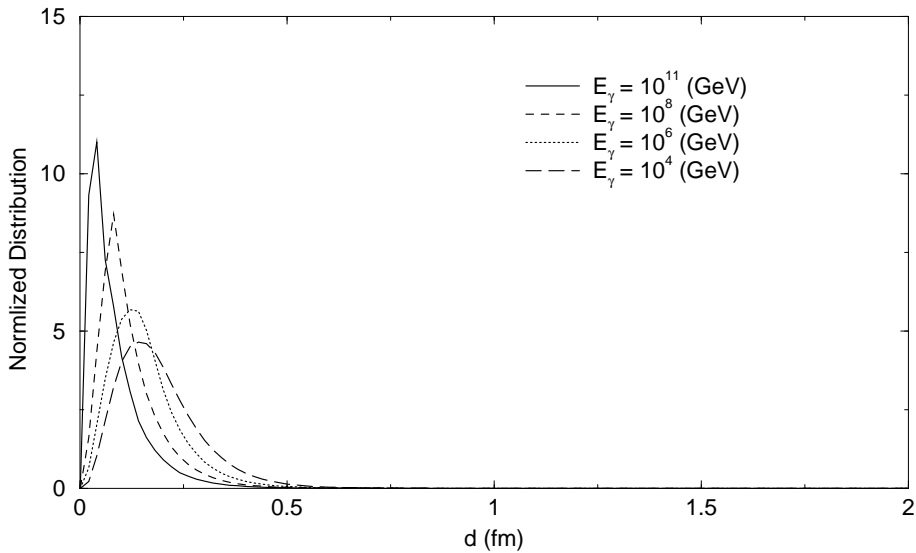}}
\caption{\label{fig:figure2b}This is the distribution of the charm contribution normalized to one.
Comparing with Fig.~\ref{fig:figure2}, we see the suppression
of massive quark contributions.  Comparing this with Fig.~\ref{fig:figure2}, we see that the distribution for
heavier quarks is more sharply peaked around small transverse sizes relative to lighter quarks.}
\end{figure}

In our calculations,
we use CTEQ5L gluon
distributions~\cite{Lai:1999wy}.  We note that, since significant sections of the matching 
region will violate the unitarity constraint, then trying to achieve an extremely
smooth interpolation is an arbitrary modification to the model which achieves
no genuine improvement.  For simplicity, therefore, we use the linear interpolation in the
MFGS model.  
The plots in Figs.~\ref{fig:figure2} and~\ref{fig:figure2b} demonstrate the resulting distribution
of hadronic sizes in the photon.  Naturally, the peak at small sizes becomes 
sharper in the UHE limit.

Before leaving the subject of $t$-dependence in the UHE photon, we recall that
in the original MFGS model of the $t$-dependence, discussed in 
Ref.~\cite{Rogers:2003vi}, the diffusion of the small size $q \bar{q}$ pairs
was neglected.  However, in a recent overview~\cite{Frankfurt:2004fm} of the
behavior of hadronic cross sections at UHE, the maximum diffusion of small size
configurations at UHE was considered.  In order to correct for the small size
diffusion, we slightly modify the diffusion factor in the amplitude of Ref.~\cite{Rogers:2003vi} 
to the form, 
\begin{equation}
F^{\pom}(t, x) = e^{\alpha^{\prime}(d) \frac{t}{2} \ln (x_{0}/x)}, \label{eq:equation10b}
\end{equation}
where $\alpha^{\prime}(d) = .25 (1 - 0.5 e^{-18 d^{4} })$~GeV$^{-2}$.  
For sizes, $d$, greater than the pion size ($d_{\pi} \approx .65$~fm), the value of $\alpha^{\prime}$ quickly
approaches the usual slope ($\alpha^{\prime} = .25$~GeV$^{-2}$) 
in accordance with the Donnachie-Landshoff soft Pomeron as in Ref.~\cite{Rogers:2003vi}.
However, unlike the Regge slope in Ref.~\cite{Rogers:2003vi} which vanishes at small sizes, the value of $\alpha^{\prime}$ 
approaches $.125$~GeV$^{-2}$ for $d \lesssim .2$~fm, which is the diffusion rate for small size dipoles
determined recently in Ref.~\cite{Frankfurt:2004fm}.  The use of the exponential function ensures that
the interpolation through the transition region is continuous, and that $\alpha^{\prime}$ approaches 
$.25$~GeV$^{-2}$ rapidly for $d \gtrsim .65$~fm and $.125$~GeV$^{-2}$ for $d \lesssim .2$~fm. 
Note that  at very high energies we should take into account that the Fourier transform of $F^{\pom}(t, x) $  
should contain a tail $\propto \exp -\mu b$ for some mass scale, $\mu$.  However, in the energy range discussed in this paper it is a small effect and hence we neglect it.

\begin{figure}
\rotatebox{270}{\includegraphics[scale=0.60]{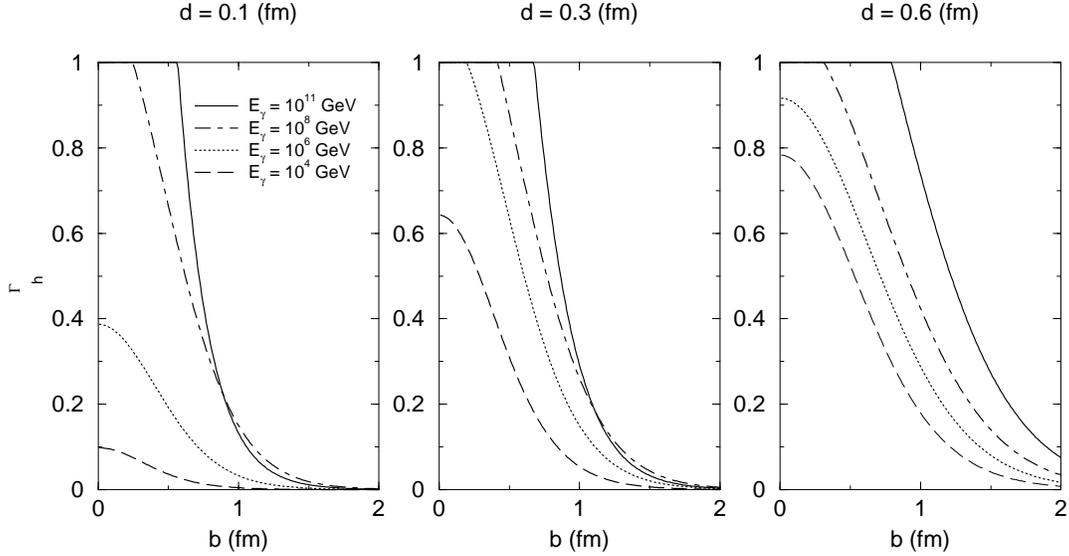}}
\caption{\label{fig:nucleon_profiles} Samples of the profile function
for the $h N$ interaction for real photon energies of $10^4$~GeV (dashed line),
$10^6$~GeV (dotted line), $10^8$ (dot-dashed), and $10^{11}$~GeV (solid line) for a range of hadronic sizes.}
\end{figure}

Samples of the profile function obtained when we use the above $t$-dependence and the
above procedure for taming the unitarity violations are shown in Fig.~\ref{fig:nucleon_profiles}.  Note the
extremely rapid growth with energy at sizes of $d = .1$~fm.
\begin{figure}
\rotatebox{270}{\includegraphics[scale=0.60]{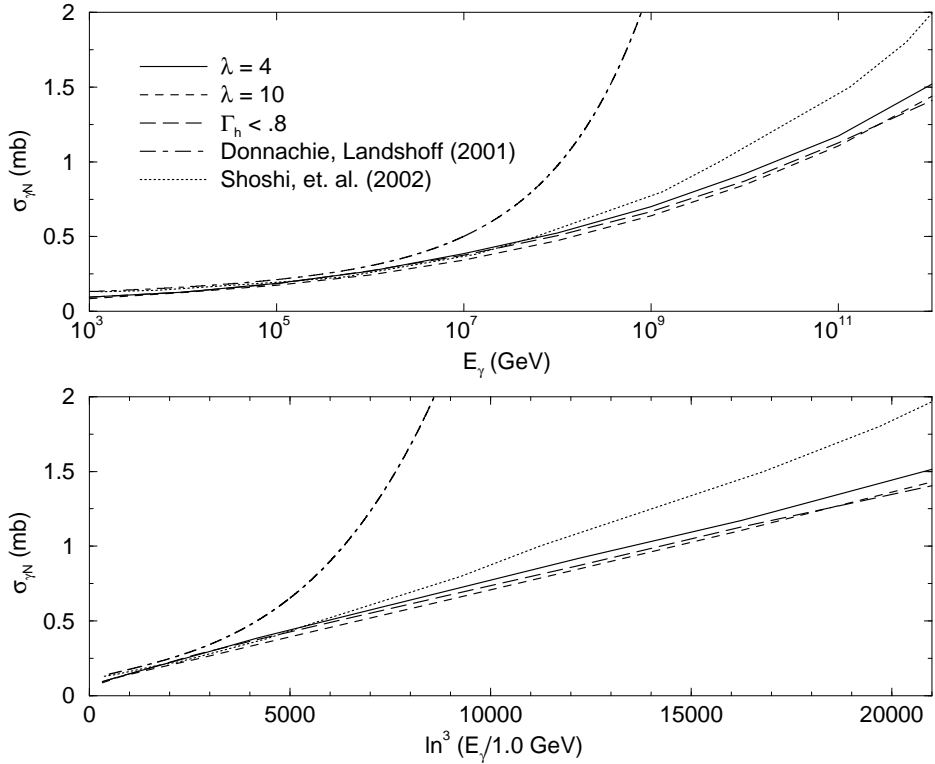}}
\caption{\label{fig:figure3}Growth of the total photon nucleon
cross section for the range of energies from $E_{\gamma}=10^{3}$~GeV to $E_{\gamma}=10^{12}$~GeV.
On the x-axis of the bottom panel, $\ln^{3}(\frac{E_{\gamma}}{E_{0}})$ ($E_{0} = 1.0$~GeV) is plotted to allow for easy comparison with 
the expected energy dependence at UHE  (see App.~\ref{sec:froissart}).  The lowest three curves in each panel are the result of using the MFGS model with 
variations in model dependent parameters to demonstrate numerical sensitivity (see text).  For comparison, the upper curve shows the two Pomeron model of 
Donnachie and Landshoff~Ref.~\cite{Donnachie:2001xx} and the dotted curve shows the model of Shoshi et. al.Ref.~\cite{Shoshi:2002in}.}
\end{figure}
\newpage
The procedure described above may be regarded as placing an upper limit on 
the growth of the cross section since we have taken the \emph{maximal} contribution that
does not violate unitarity.    
The resulting nucleon cross section has been plotted in Fig.~\ref{fig:figure3}.
To test the numerical sensitivity to a variation in the upper limit of the profile function, we have included
the result of placing the upper limit of the profile function at $0.8$ rather than
at $1.0$.  
In the studies ~\cite{Frankfurt:1995jw, McDermott:1999fa} 
 the matching parameter, $\lambda$ was estimated based on the analysis 
 of the expressions for $\sigma_L(x,Q^2)$ to be of order 10. A later analysis of the $J/\psi$ production \cite{Frankfurt:2000ez}
 suggested that a better description of the
 cross section for the intermediate $0.5\ge d\ge 0.3 fm$ is given
 by $\lambda \sim 4$ while the
  cross section in the perturbative region depends very weakly
on  $\lambda$. We therefore include in Fig.~\ref{fig:figure3} the result of using $\lambda = 10$ to test the
sensitivity to the matching ansatz.  It is evident that variations in these parameters make a small reduction in the upper limit.  For the
rest of this paper, we assume that $\lambda = 4$ and that the unitarity constraint implies, $\Gamma_{h} \leq 1$.

For comparison, the Donnachie-Landshoff two-Pomeron model~\cite{Donnachie:2001xx} with no unitarity constraint is also shown in Fig.~\ref{fig:figure3}. 
Yet another approach was used for calculations in Ref.~\cite{Shoshi:2002in}.  
The result of this 
calculation is shown as the dotted curve in Fig.~\ref{fig:figure3}.    
In the model of Ref.~\cite{Shoshi:2002in}, the large size components are modeled by non-perturbative 
QCD techniques, but the small size components use a Donnachie-Landshoff two Pomeron model
with parameters somewhat different from the original model.
  While the method used in Ref.~\cite{Shoshi:2002in}
imposes impact parameter space unitarity in the dipole picture, it 
does not account for the non-normalizability of the photon wavefunction.  Instead, a unitarity constraint is imposed upon the
$S$-matrix for $\gamma$-proton scattering, and is thus less restrictive than our approach which applies unitarity constraints to the profile function for
individual hadronic configurations.   Our approach thus restricts the growth of the cross section more than either of the two above scenarios as can 
be seen in Fig.~\ref{fig:figure3}.

\begin{figure}
\rotatebox{270}{\includegraphics[scale=0.40]{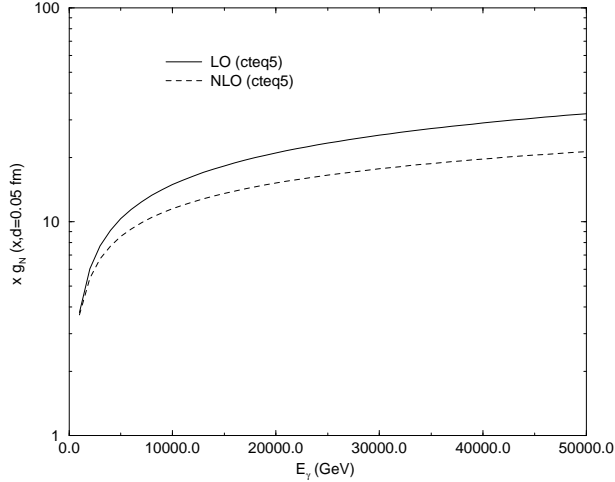}}
\caption{\label{fig:pdfcomp}Plot comparing the the growth of LO and 
NLO pdfs at high energy for the small dipole size, $d = .05$~fm.}
\end{figure}

It has been observed that agreement with data is improved if
one uses MS-bar NLO pdfs in very low-x DIS experiments.  Our main
concern is that the rate of increase described by the interpolation 
that we obtained in Eqs.~\ref{eq:param1}-\ref{eq:param3}
of the appendix is not drastically modified by the inclusion of higher order
effects.  In particular, the \emph{rate} of growth, and the order of 
magnitude of higher order corrections should not be drastically 
altered from what one obtains at leading order.
To argue that this is the case, we notice that the
LO and NLO parton densities grow at nearly the same rate at very
high energies.  This is demonstrated for a typical small 
configuration size in Fig.~\ref{fig:pdfcomp} where we compare the LO and NLO CTEQ5 parton
densities lowest values of $x$ where parameterizations exist.  The main 
difference between the two parton densities is that, in the
high energy limit, the leading order gluon density is nearly
a constant factor larger than the NLO pdf.  (This makes sense
when one considers the color factor 9/4 difference between
quark-antiquark dipoles, and gluon dipoles 
and the need to include $q\bar q g$ dipoles in a consistent NLO formulation of the dipole model.)
Indeed, above about $E_{\gamma} = 20000$~GeV in the plot of Fig.~\ref{fig:pdfcomp},  the LO curve is shifted by
about a constant factor of $9/4$ upward from the NLO curve, and the energy dependence used (see Eq.~(\ref{eq:param2}) of App.~\ref{sec:smallx}) 
describes both curves with an accuracy of $\le 10$\% for the energy range $E_{\gamma} > 20,000$~GeV.  In the spirit of obtaining an \emph{upper limit} on the growth of the 
cross section, we continue to used the LO pQCD result.
 Since we checked that our LO inspired parameterization of $\sigma(d,x')$ describes the data well down to $x\sim 10^{-3}$ the difference in the normalization is likely to be partially absorbed in the definition of the cross section. Note also that the recent studies of the small x behavior of the
the gluon densities indicate that the NLO approximation is close to the resumed result down to $x\sim 10^{-7}$, see a review in \cite{G.Salam}.

\begin{figure}
\rotatebox{270}{\includegraphics[scale=0.40]{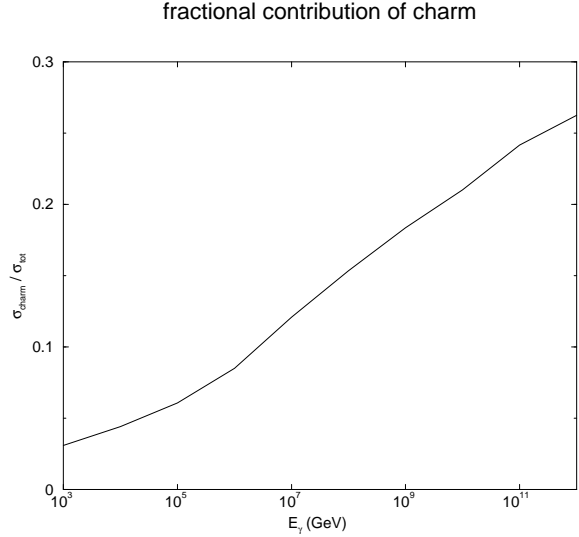}}
\caption{\label{fig:charm}The fraction of the total hadronic $\gamma$-proton cross section due to charm $q \bar{q}$ pairs.}
\end{figure}

Before leaving the discussion of the nucleon target, we note
several advantages and disadvantages of the current approach:
First, we note that Fig.~\ref{fig:figure3} is consistent with the rate growth of the cross
section for a real photon at UHE  which we find in App.~\ref{sec:froissart} to be $\sim \ln^{3} (E_{\gamma}/ E_{0})$, where $E_{0}$ is of
order $1$~GeV (see e.g.~\cite{Donnachie:en} for a review of the Froissart bound for the hadron - hadron scattering cross section, See App.~\ref{sec:froissart} for
a generalized discussion of energy dependence for real photons.)  
We are also able to analyze the contribution from different flavors separately.
In Fig.~\ref{fig:charm}, we see that the fraction of the total cross section due to charmed particles rises to nearly 
$25\%$ at $10^{12}$~GeV~\footnote{This calculation of the fractional contribution of charm assumes that the cross section for the scattering of each quark flavor individually grows at the
maximum rate allowed by unitarity.  The fractional contribution of charm could, of course, be much larger (and hence a more complete restoration of flavor SU(4)) 
if there is some mechanism that suppresses 
the light quark contribution but not the charm contribution.}.  This indicates that a large fraction of showers initiated 
by UHE photons should contain a pair of leading 
charmed particles. 
It is worth investigating whether such showers have a substantially longer penetration depth in the atmosphere and could be separated by 
the Auger detector.

\begin{figure}
\rotatebox{270}{\includegraphics[scale=0.40]{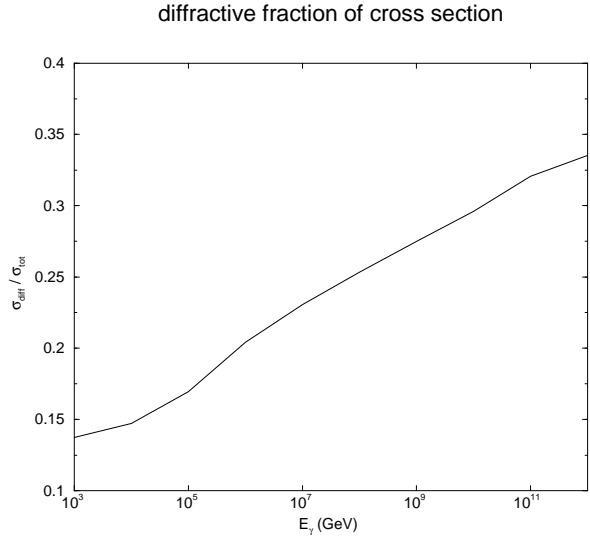}}
\caption{\label{fig:diffractive}The fraction of the total hadronic $\gamma$-proton cross section due to diffractive reactions of the hadronic component.
The large fraction of the cross section due to diffractive scattering at $E_{\gamma} = 10^{12}$~GeV indicates relative nearness to the black disk limit.}
\end{figure}

Furthermore, by recalling the relationship between  the diffractive component of the basic cross section, $\hat{\sigma}_{diff}$ and the hadronic profile function,
\begin{equation}
\hat{\sigma}_{diff} = \int d^{2} {\bf b} \left| \Gamma_{h}({\bf b}) \right|^{2}, 
\end{equation}
we may also separate the fraction of the total hadronic profile function due to diffractive scattering.  
This is shown in Fig.~\ref{fig:diffractive}, which gives a integrated measure of the proximity to the black disk limit.
In the black disk limit, diffractive scattering accounts for exactly half of the total cross section.  Therefore, the fact that,
as Fig.~\ref{fig:diffractive} shows, the fraction of the cross section due to diffractive scattering is around .35 at UHE indicates
that diffractive scattering plays a significant role and that there will be large shadowing in nuclei.

Since the cross section grows extremely
quickly at UHE, then the unitarity limit is saturated even at values of
dipole size around  $d = 0.1$~fm.  Figure~\ref{fig:figure2}
demonstrates that the largest contributions to the total
photo-production cross section come from regions around $d \approx .1$~fm
and from the transition region.  Sizes smaller than this contribute
very little to the total cross section.  Thus, pQCD 
provide very little detailed information, since all models of the basic cross section
(at small sizes) which 
violate the unitarity constraint at $d \approx 0.1$~fm will give very
similar results.  
Furthermore, since the gluon distribution rises very
sharply between $d = 0$~fm and $d = 0.1$~fm at very high energies, then the calculation in
Eq.~(\ref{eq:equation1}) becomes  more sensitive to how the gluon
distribution is sampled.  Thus, there is more sensitivity to the
parameter, $\lambda$, used to relate the hadronic size to the hardness
of the interaction.   As seen in Fig.~\ref{fig:figure3}, if $\lambda=10$, then
the cross section is suppressed by around ten percent from its value when $\lambda = 4$.

The largest source of theoretical uncertainty comes from the region of
large hadronic sizes.  Though the behavior of large configurations can
be reasonably expected to follow pion behavior at accelerator
energies, we do not have any experimental data for $ \gtrsim
10^{6}$~GeV hadrons with which to model these extremely high energy Fock
states.  Moreover, we have so far been associating each hadronic Fock
state with a particular size.  It may be that as the energy of the
photon increases, a large number of hadronic Fock states (perhaps
multiple pion states) may be associated with a single size.  Moreover,
contributions from large impact parameters become significant for
extremely energetic photons.  Thus, predictions become sensitive to
how the model handles the $t$-dependence of the amplitude.  The current model
is based on the assumption that the typical $t$-dependence for low
energies continues into the UHE regime.  Note, however, that in the case of the $pp$ scattering the analysis of \cite{Frankfurt:2004fm}
indicates that, though the black disk limit leads to the slope $B\propto \ln^2 s$, this is a very small correction for the energies discussed here.
Finally, we want to emphasize that we allow the impact parameter amplitude to approach $\Gamma = 1$ without a slowdown at $\Gamma \sim 0.5$ as happens 
in many other dipole models where eikonal type parameterizations of the dipole-nucleon cross section are used (see, for example,~\cite{Bartels:2002cj}).  
One again it is consistent with our aim of yielding a conservative upper bound.
\section{nuclear targets}
\label{sec:section3}
Since we are interested in the interaction of UHE photons with atmospheric nuclear 
targets, we now go on to investigate the growth of the UHE cross section for the case 
of a real photon scattering from a nuclear target. 
The steps apply to any of light nuclei constituting the atmosphere, but we use
$^{12}$C for the purpose of demonstration since the $^{12}$C nucleus has the 
approximate number of nucleons for a typical atmospheric nucleus.

\begin{figure}
{\includegraphics[scale=0.50]{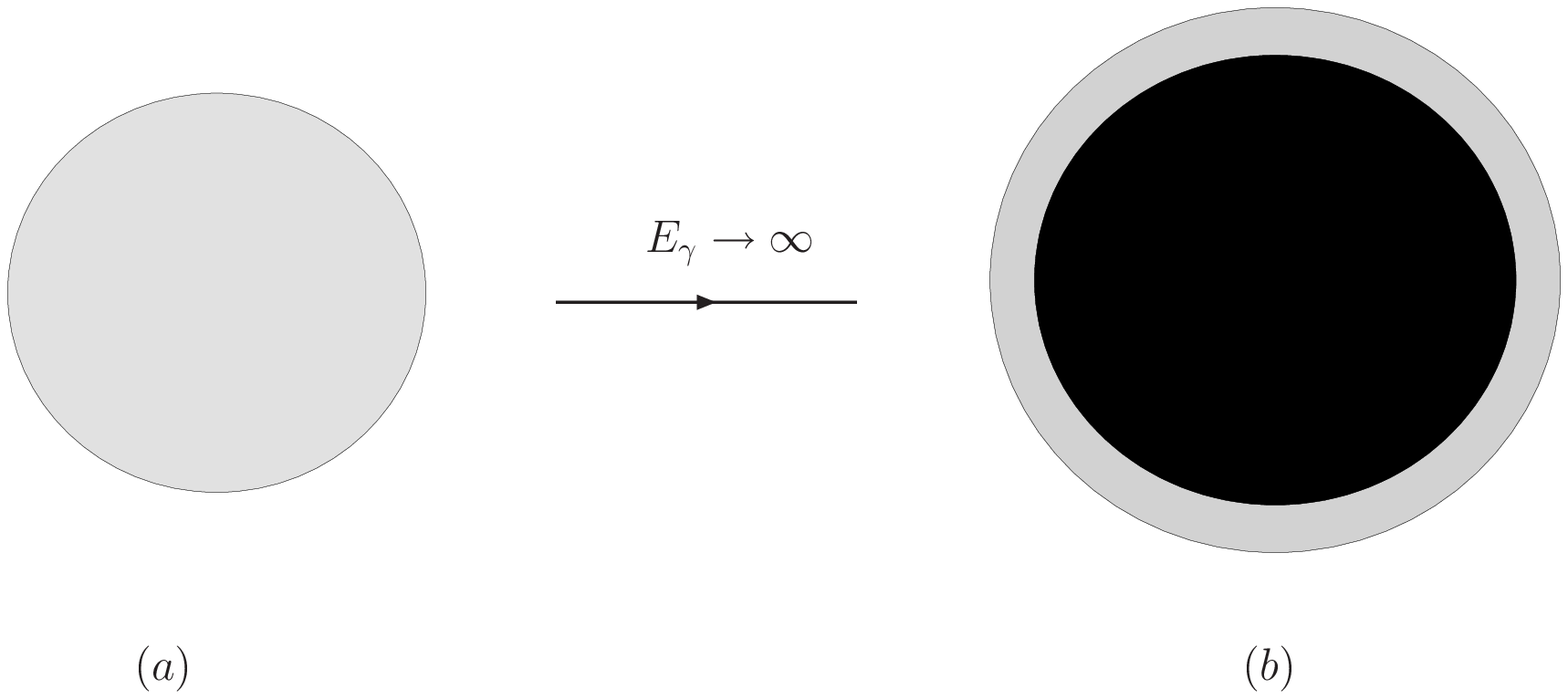}}
\caption{\label{antishad2}The nuclear disk as ``seen'' by the incident hadronic 
configuration in the simple homogeneous model of the unintegrated nuclear PDFs.  The 
level of absorption by the disk is indicated by the level of grayness.  At low energies, (a), the
disk is weakly absorbing and homogeneous.  At very high energies in (b) 
the disk becomes black and is thus totally absorbing.\\}
{\includegraphics[scale=0.50]{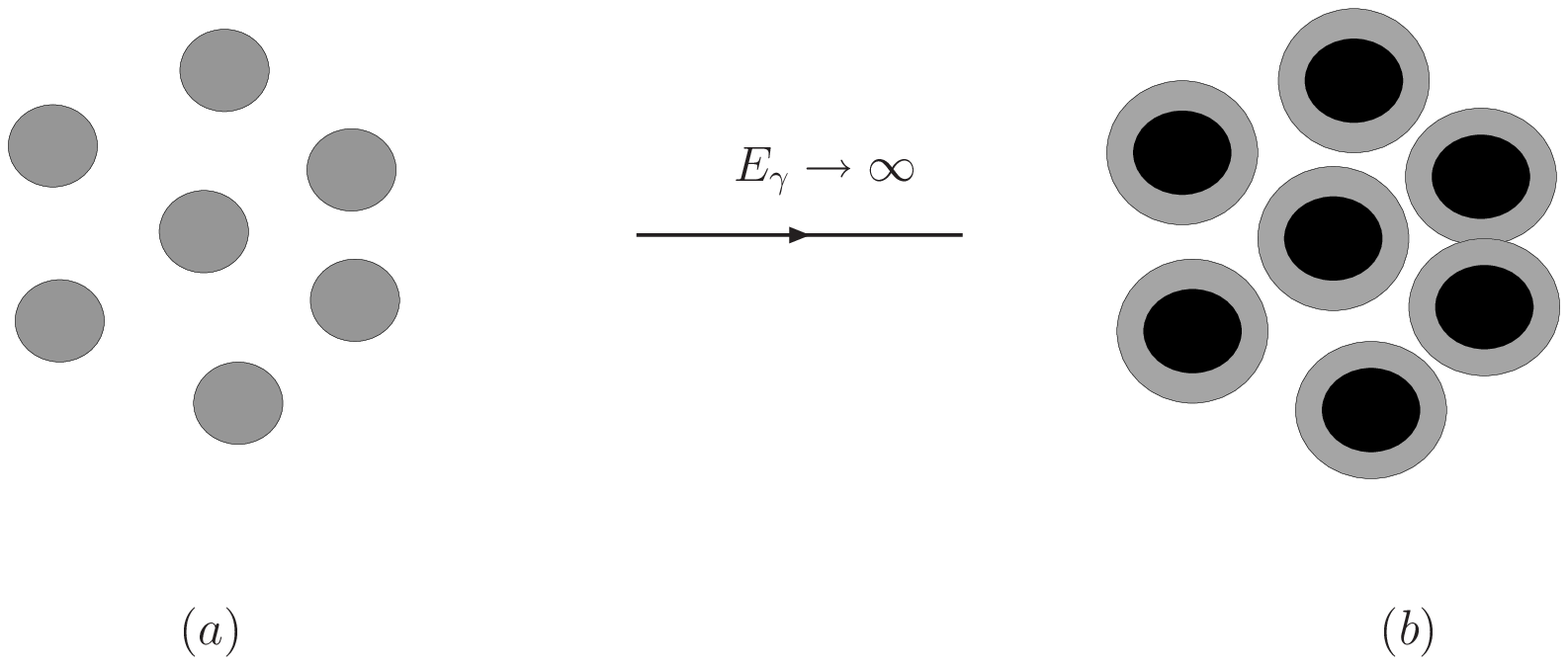}}
\caption{\label{antishad} A more accurate way to visualize what the incident hadronic configuration probably
actually ``sees''.  The nucleus consists nucleons separated over a large distance with concentrated nuclear matter in (a). 
At very high energies (b), each of the nucleons becomes totally absorbing (black).}
\end{figure}

At first glance it seems natural to repeat the procedure we followed for the proton case in Sec.~\ref{sec:section2} with the profile function given by a nuclear shape, and cross section for
the interaction of the small dipoles as given by Eq.~(\ref{eq:pertex}).
However we found that if we follow this procedure we end up with the
obviously wrong result that in  
 the UHE limit, this approach quickly leads to the situation that
$\sigma_{\gamma A} >> A \sigma_{\gamma N}$ which is physically unreasonable.
The reason for this becomes clear if we visualize the 
relationship between the proton PDF and the nuclear PDF.  
The unitarity condition is not sensitive to  effects of transverse correlations of the partons. 
The unitarity constraint would tame the dynamics if one could assume  that the nucleus is a
perfectly homogeneous distribution of nuclear matter.  At high energies, the disk of nuclear
matter ``seen'' by the incident hadronic configuration blackens as represented 
schematically in Fig.~\ref{antishad2}.  Any inhomogeneity in the distribution
of nucleons is accounted for at low energies in the grayness of the nuclear disk without
yielding a quantitative difference in the total cross section.  However, the actual 
distribution of nuclear matter in light nuclei ``seen'' by the incident dipole probably looks more like that 
of Fig.~\ref{antishad} - a collection concentrated regions of nuclear matter which individually 
grow black in the high energy 
limit but far apart from each other transversely. 
If the nuclear system is dilute and nucleons do not overlap transversely, the use of the $\Gamma(b) < 1$ condition becomes insufficient.  
Thus, we will certainly find that $\sigma_{\gamma A} >> A \sigma_{\gamma N}$
 if we assume that both 
the disk of the individual nucleon \emph{and} the disk of the nucleus grow black in 
the high energy limit.  That is, the cross section resulting from Fig.~\ref{antishad}(b) is 
certainly larger than the sum of the cross sections from each of the 
blackened nucleons seen in Fig.~\ref{antishad2}(b).
The simplest illustrative example would be to consider scattering off the deuteron - in this case neglect of the cluster structure of the system would grossly overestimate the maximal cross section 
for the interaction of this system with a small dipole.
It is worth emphasizing that  all these considerations are valid for the light nuclei and are likely to be a correction for the scattering off sufficiently heavy nuclei.
A more meaningful upper estimate of the cross section of scattering off light nuclei is, therefore, to take into account
\emph{first} the taming of the elementary cross sections and next 
the Glauber - Gribov theory of nuclear shadowing due to diffraction
~\cite{Glauber, Gribov}
which does not rely on the twist decomposition of the cross section. 
Since we observed that diffraction constitutes a large fraction of the total cross section we expect that a large shadowing effect will emerge with 
growing energy. 
Consequently, our result will automatically be consistent with S-channel unitarity.
Hence, we  
take the usual approach to nuclear scattering when the product of the 
nuclear optical density with the cross section is small: $A \sigma_{\gamma N} T(b) < 1$.     
Following in the spirit of the treatment of hadronic fluctuations in the Good and Walker picture 
as presented in e.g.~\cite{Blaettel:1993ah}, we consider the states of the incident photon to be a linear 
combination of nearly ``frozen'' hadronic states that do not mix with one another during 
their passage through the target.  Each of these states scatters from the target with a cross
section $\langle \hat{\sigma}_{h} \rangle = \hat{\sigma}_{h}$.  
Let us write the total photon-nucleus cross section as, 
\begin{equation}
\sigma^{\gamma A} = A \hat{\sigma}_{h} - \Delta \sigma.
\end{equation}
A standard result of the Glauber-Gribov theory in the language of hadronic fluctuations is that the full shadowing correction, $\Delta \sigma$, can be written as,
\begin{equation}
\label{eq:shadapprox}
\Delta \sigma = \frac{A}{4} \int d^{2}{\bf b} \, T^{2}(b) \langle \hat{\sigma}_{h}^{2} \rangle e^{-\frac{1}{2} A \langle \hat{\sigma}_{h} \rangle T(b)}. 
\end{equation} 
Where $T(b)$ is the nuclear optical density normalized to unity.
Equation~(\ref{eq:shadapprox}) is an approximate formula valid for the case of small fluctuations or small nuclear thickness (the later is true in our case). 
We will find a convenient expression for $A_{eff}/A$ if we expand this expression using the argument of the exponent as a small parameter:
\begin{equation}
\label{eq:nucexpand}
\sigma^{\gamma A} = A \hat{\sigma}_{h} - \frac{A \langle \hat{\sigma}_{h}^{2} \rangle}{4} \int d^{2}{\bf b} \, T^{2}(b) + \cdots
\end{equation}
If $A_{eff}$ is defined by the relation, $\sigma^{\gamma A} = A_{eff} \sigma^{\gamma N}$, then dividing Eq.~(\ref{eq:nucexpand}) by $A \sigma^{\gamma N}$ gives,
\begin{equation}
\frac{A_{eff}}{A} = 1 - \frac{\langle \hat{\sigma}_{h}^{2} \rangle}{4 \sigma^{\gamma N}} \int d^{2}{\bf b} \, T^{2}(b) + \cdots.
\end{equation}
Define an effective cross section,
\begin{equation}
\sigma_{eff} \equiv \frac{ \langle \hat{\sigma}^{2} \rangle}{\sigma^{\gamma N}} = \left.  \frac{16 \pi}{\sigma^{\gamma N}} \frac{d \hat{\sigma}_{h}^{diff}}{dt} \right|_{t = 0},
\end{equation}
where by definition,
\begin{equation}
\langle \hat{\sigma}^{2} \rangle = \int_{0}^{1} dz \int d^{2} {\bf d} \left| \psi_{T}(z,d)
\right|^{2} \hat{\sigma}^{2}_{h}(d,x^{\prime})\,.
\end{equation}
Then we can write,
\begin{equation}
\label{eq:aexpand}
\begin{split}
\frac{A_{eff}}{A} = & \frac{\sigma_{eff} - \frac{\sigma_{eff}^{2}}{4} \int d^{2}{\bf b} \, T^{2}(b) + \cdots}{\sigma_{eff}} \\
                 & =  \frac{\int d^{2}{\bf b} \, \left( T(b) \sigma_{eff} - \frac{\sigma_{eff}^{2}}{4} T^{2}(b) + \cdots \right)}{\sigma_{eff}}.
\end{split}
\end{equation}
Considering the first two terms in Eq.~(\ref{eq:aexpand}) as the first terms in a power series expansion\footnote{In principle one can write a more accurate formula which would take into account
deviations of $\frac{ \langle \hat{\sigma}^{n} \rangle}{\sigma_{\gamma N}}$ from 
$\sigma_{eff}^{n-1}$ for $n\ge 3$. However, numerically these effects are small especially for the light nuclei where double scattering gives a dominant contribution.  For all of our calculations, 
$\sigma_{eff} T(b)$ is small enough to justify an expansion.}, we may write, 
\begin{equation}
\frac{A_{eff}}{A} = \frac{2 \int d^{2} b \, (1 - e^{-\frac{1}{2} \sigma_{eff} T(b)})}{\sigma_{eff}},
\label{eq:shadratio} 
\end{equation}
If we identify the effective profile function as,
\begin{equation}
\Gamma_{eff}(b) \equiv 1 - e^{-\frac{1}{2} \sigma_{eff} T(b)},
\label{eq:nuclearprofile}
\end{equation}
then we see that both the effective profile function and the shadowing ratio are less than
unity by construction.  Furthermore, because $A_{eff}$ must be less than $A$ (for scattering in any range of 
impact parameters), then we may regard $\Gamma_{eff}(b) < 1$ as our unitarity condition for the nucleus.

(Note that the diffractive components have mass squared proportional to $1/d^{2}$ and therefore correspond to 
rather 
large hadron multiplicities.)
The result of evaluating Eq.~(\ref{eq:shadratio}) and solving for the $\gamma A$ cross section
is shown in Fig.~\ref{fig:nucldist}.
\begin{figure}
\rotatebox{270}{\includegraphics[scale=0.40]{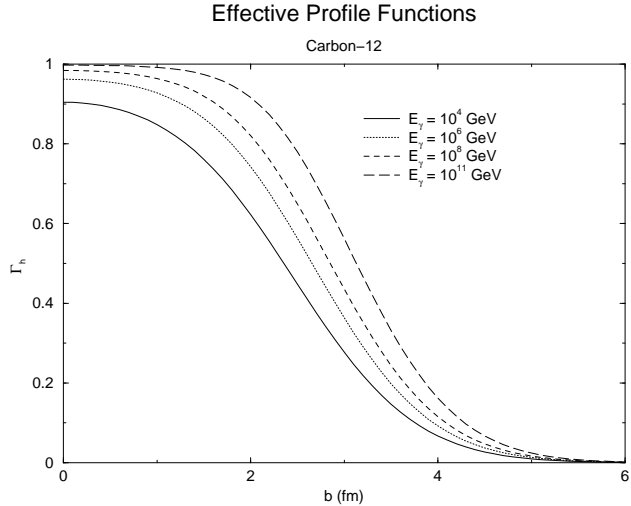}}
\caption{\label{fig:nuclprof}Sample of the effective profile functions (see Eq.~(\ref{eq:nuclearprofile})) for a real photon 
on a $^{12}$C target.}
\end{figure}
\begin{figure}
\rotatebox{270}{\includegraphics[scale=0.60]{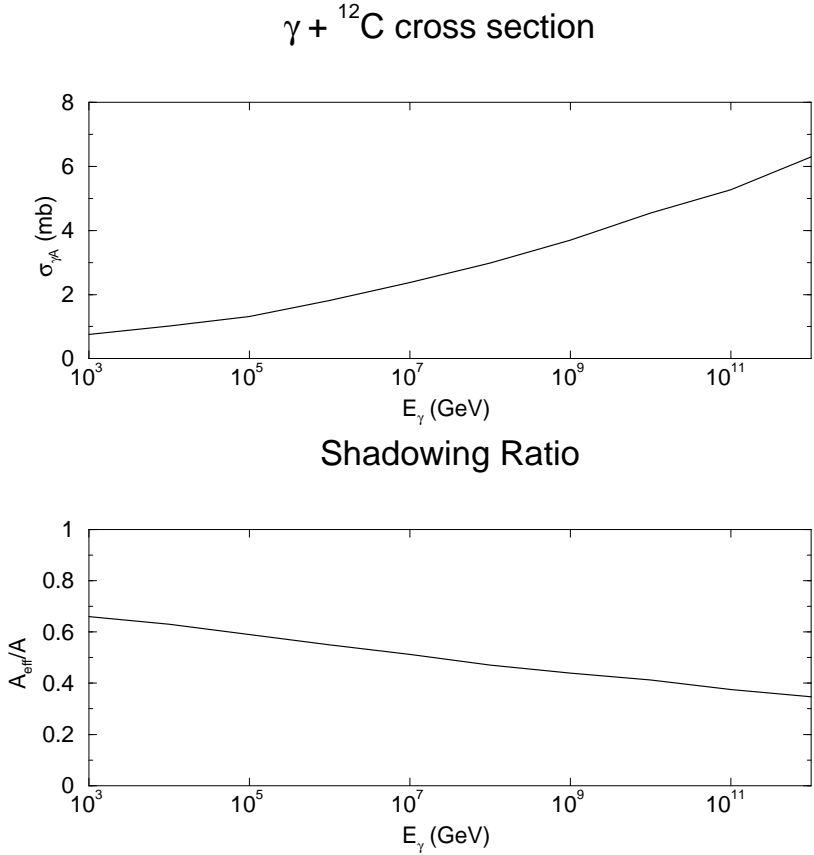}}
\caption{\label{fig:nucldist}The upper panel shows the dependence of the $\gamma$-$^{12}C$ cross section on the incident 
photon energy.  The lower panel shows the dependence of the shadowing ration $\sigma_{\gamma A}/(A \sigma_{\gamma N})$.}
\end{figure}
\begin{figure}
\rotatebox{270}{\includegraphics[scale=0.40]{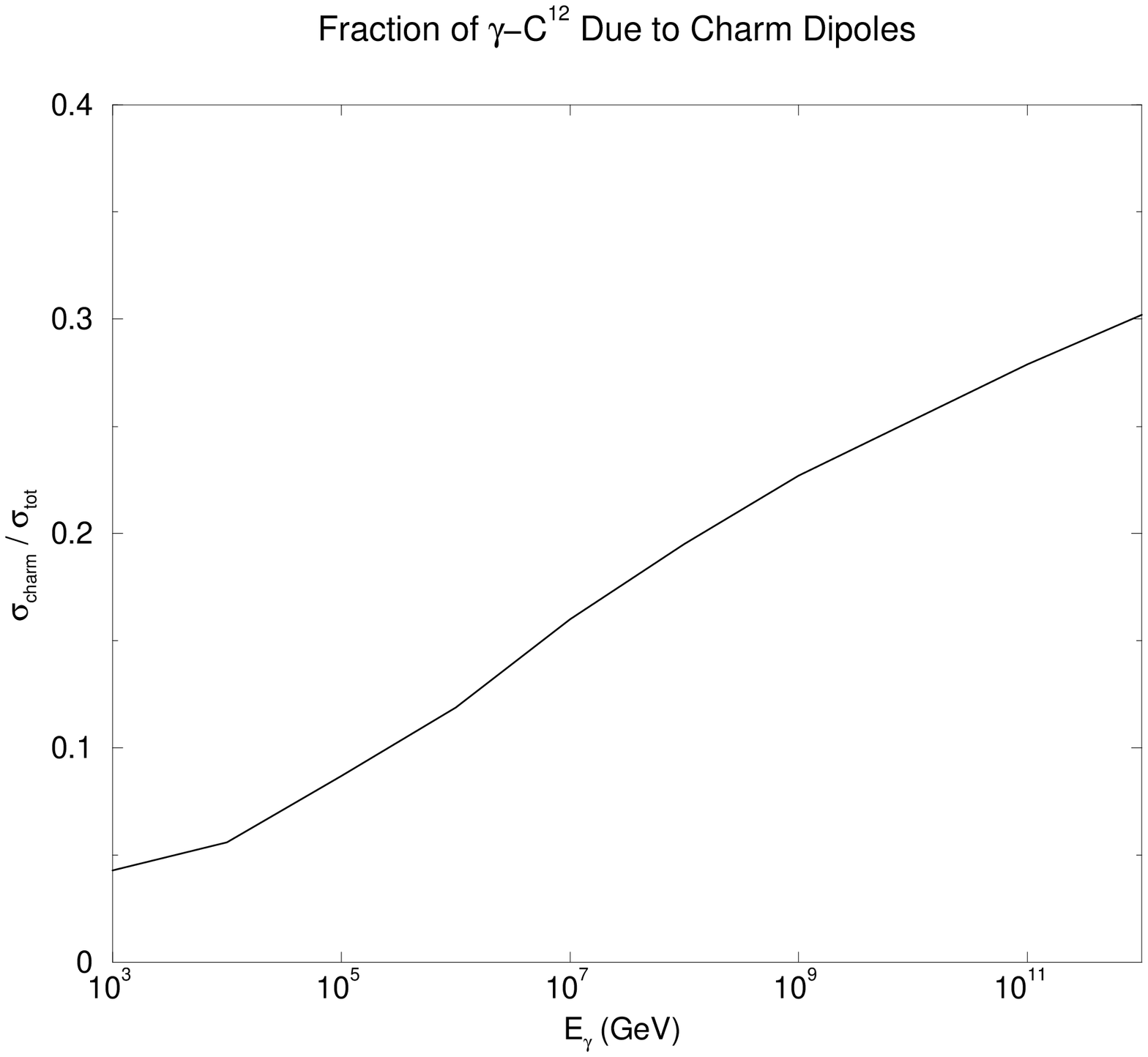}}
\caption{\label{fig:carbcharmratio}The fraction of the total $\gamma$-$^{12}C$ cross section due to charm dipoles.}
\end{figure}
One can see that the energy dependence of the $\gamma$-$^{12}$C scattering cross section is substantially weaker than for the proton target due to 
nuclear shadowing, though the increase of the cross section as compared to the energies studied experimentally is still large.  Since there is less 
shadowing for the case of incident charm dipoles, then our analysis indicates that there is a larger fraction of the total photon-$^{12}$C cross section that arises due to charm 
dipoles than in the case of a proton target.  This is shown in Fig.~\ref{fig:carbcharmratio}.  
In the case of a target $^{12}$C nucleus, the fraction of the total cross section due to charm dipoles is around $30\%$.
The large amount of shadowing that we find already has implications for energies around $\sim 100$~TeV which are 
relevant to a number of current cosmic ray experiments~\cite{TeV}.  In addition, the forthcoming studies of ultra-peripheral 
heavy ion collisions at the LHC would allow, to some extent, a check of our predictions by measuring shadowing for photon-heavy ion 
interactions for the range of values for $\sqrt{s}$ from $1000$~GeV to $2000$~GeV.  Within our model, we find that $A_{eff}/A \approx .3$ for $\sqrt{s} = 100$~GeV
and $A_{eff}/A \approx .2$ for $\sqrt{s} = 2000$~GeV with $A = 220$.  
In order to allow for a simple extrapolation from the shadowing ratio, $(A_{eff}/A)_{C}$, for Carbon to other nuclei with masses typical of atmospheric 
atoms we use the function,
\begin{equation}
\frac{A_{eff}}{A} = \left( \frac{A_{eff}}{A} \right)_{C} \left( \frac{A}{12} \right)^{n}, 
\end{equation}
where we then determine $n$ for a a set of fixed photon energies and for the range of atomic masses, $12 \leq A \leq 16$.
As a sample, we list the following: For $E_{\gamma} = 10^{12}$~GeV, we find $n = -.41$; for $E_{\gamma} = 10^{9}$~GeV, we find $n = -.35$; and for $E_{\gamma} = 10^{6}$~GeV, we find $n = -.3$.

\section{Conclusion}

In this paper we have investigated the photon-proton/nucleus cross section 
in the range of energies from $10^{3}$~GeV to $10^{12}$~GeV.
Figure~\ref{fig:figure3} demonstrates that the total cross section
rises by about a factor of 12, but that there is a significant amount of
uncertainty involved when the details of the model are varied.  This result, however,
gives us a very reasonable upper limit on the cross section since we have consistently
taken the maximum cross section allowed by unitarity.  
The cross section
varies approximately linearly with
$\ln^{3}(\frac{E_{\gamma}}{E_{0}})$  (see App.~\ref{sec:froissart}).  (We use $E_{0} = 1.0$~GeV which is
consistent with hadronic sizes; see, for example, page 18 of
Ref.~\cite{Donnachie:en}).  

The solid curve 
in Fig.~\ref{fig:figure3} provides a reasonable estimate (or if we prefer to take a more cautious attitude, an upper limit) to the 
$\gamma N$ cross-section at extremely high photon energies.  We note that the dipole approach is 
consistent with the direct extrapolation of the photo-nuclear cross section~\cite{Bezrukov:1981ci} and with 
a model based on unitarity in the $t$-channel~\cite{Cudell:2002ej}, but not with the model of~\cite{Donnachie:2001xx} which 
uses a two-Pomeron approach and places no unitarity constraint on the growth of the cross section.  
There is also disagreement
with the model of Ref.~\cite{Shoshi:2002in} which uses the Donnachie-Landshoff two-Pomeron approach, but applies a unitarity
constraint to the $S$-matrix for $\gamma$-proton scattering rather than to the profile function of individual hadronic components.

We have shown in Fig.~\ref{fig:charm} that as much as $25 \%$ of the cross section may be due to charmed mesons for the case of 
a target proton, and this fraction rises to $30 \%$ when we consider a target $^{12}$C nucleus.
For the $^{12}$C target, we use a direct application of the usual Gribov-Glauber theory, and we find 
the there is a
large amount of shadowing increasing with energy.  It is worth noting here that though our results for the elementary $\gamma p$ cross section are rather close to the results of 
\cite{Engel:1996yb} for $\sqrt{s}\le 10^3 \, GeV$ considered in this paper using a 
generalized vector dominance model with a point-like component in the photon wave function, the model of Ref.~\cite{Engel:1996yb} leads to a 
nuclear shadowing effect which is practically energy independent.

This is consistent with 
the dipole-proton interaction approach to the black disk limit.
The resulting cross section is shown in Fig.~\ref{fig:nucldist} which indicates a rise in the cross section 
of about a factor of 7 when the energy increases from $10^{3}$~GeV to $10^{12}$~GeV.
The relevance of these observations is in the characterization of atmospheric showers
induced by UHE neutrinos and super-GZK cosmic rays, where upper limits on the allowed growth of the 
photon-nucleus cross section are needed.

\begin{acknowledgments}
We would like to thank Vadim Guzey for discussions on the use of
the nuclear parton distribution functions.  Much of the introductory material related to cosmic ray implications arose directly from 
discussions with Spencer Klein and Venya Berezinsky and we thank them for their valuable input.  We would also like to thank Ralph Engel and
Markus Risse for very valuable discussions.  This work is supported by DOE grants under contract DE-FG02-01ER-41172
and DE-FG02-93ER40771.
\end{acknowledgments}

\appendix
\section{Interpolation To Very Small Bjorken-x}
\label{sec:smallx}
In order to further extrapolate the
basic, small size, cross section to extremely small values of $x$, we
make a fit to the CTEQ5L gluon distribution in the region of lowest 
$x$ ($10^{-5} > x > 10^{-4}$) where the parameterizations exist.  We find that the following interpolation
agrees to within a few percent over the range of small-$x$ and for $.01$~fm $ > d > .2$~fm:
\begin{eqnarray}
x g_{N}(x,d) = a(d) x^{c(d)} \label{eq:param1} \\
c(d) = -.28 d^{-.11}  \label{eq:param2} \\
a(d) = 3.9 - 13.9 d + 20.1 d^{2} + .5 \ln d. \label{eq:param3}
\end{eqnarray}
\section{Limits on Energy Dependence at Ultra-High Energies.}
\label{sec:froissart}
It is possible to obtain the general energy dependence, $\sigma^{\gamma h} \sim \ln^{3} E_{\gamma}$ for the $\gamma$-hadron cross section in the UHE limit within 
the dipole model.  Here we give a general proof 
based only on the following assumptions about the UHE limit:
\begin{itemize}
\item The dipole model for a finite number of active quark flavors holds for the real photon in the UHE limit in the sense that
at a fixed energy, the dipole cross section increases with $d$ no faster than $d^{2}$.
\item A given finite size hadronic Fock component of the real photon scatters with exactly the maximum possible cross section allowed
by the unitarity constraint when $E_{\gamma} \rightarrow \infty$, and the rate of increase
of the cross section for each individual Fock component of size, $d$, is limited by the rate of growth in the Froissart limit.
\item The very small size hadronic Fock components of the real photon scatter with a cross section whose rate of growth is 
no faster than a power of $x^{\prime}$ (or $E_{\gamma}$).  
\end{itemize}
The second bullet above requires some clarification.  Usually, the Froissart bound is only applied to the interaction
between two hadrons rather than to the interaction between a wave-packet and a hadron.  
However, since the derivation of the Froissart bound is based on analyticity in the $t$-channel which leads to the 
requirement that the amplitude in impact parameter space falls of at least as fast as $e^{-2 m_{\pi} b}$~\cite{Donnachie:en} then the argument
works in our case as well~\footnote{Actually, in the MFGS model, the amplitude behaves as $\sim e^{a t \ln \frac{x^{\prime}_{0}}{x^{\prime}}}$ in the UHE limit where $a$ is a positive constant.  So, it falls off
\emph{faster} than $e^{-2 m_{\pi} b}$ in impact parameter space and therefore the cross section should not increase at precisely the maximum rate allowed by unitarity.  
Therefore, the fact that the MFGS model is nearly linear in $\ln^{3} \frac{E_{\gamma}}{E_{0}}$ in Fig.~\ref{fig:figure3} indicates that
pre-asymptotic affects are still significant in the considered energy range.}.

We are only interested in the variation of the cross section with energy.  Therefore we will leave out over-all 
factors in order to simplify the argument.  Note that $\ln \frac{x^{\prime}}{x_{0}^{\prime}}$ can always be separated into a sum of $\ln (E_{\gamma})$ and terms 
that only depend on $d$.  For the rest of this section, we will always write $\ln \frac{x^{\prime}}{x_{0}^{\prime}}$ as $\ln(E_{\gamma})$ since it is the leading
powers of photon energy that will interest us.
To be concise, the symbol $\sim$ will indicate how a cross section varies with photon energy, $E_{\gamma}$, whereas 
$\propto$ will indicate how a cross section varies with hadronic size, $d$.  The first of the above assumptions
allows us to state that,
\begin{equation}
\sigma(E_{\gamma}) \sim \sum_{flavor} \, \int_{0}^{\infty} d \, dd \left| \psi(d)
\right|^{2} \hat{\sigma}(d,x^{\prime})\,.
\label{eq:proof1}
\end{equation}
The integral over momentum fraction from Eq.~(\ref{eq:equation10}) is assumed to be implicit.
Also, for the rest of this section, the sum over flavors in Eq.~(\ref{eq:proof1}) will be understood and left out.
Since the energy dependence of the integrand in Eq.~(\ref{eq:proof1}) can 
be understood in the extreme limits of $d \rightarrow 0$ or $d \rightarrow \infty$, but is model dependent
in the intermediate range of $d$, then let us separate Eq.~(\ref{eq:proof1}) into the sum of three terms:
\begin{equation}
\sigma(E_{\gamma}) \sim \stackrel{region 1}{\overbrace{\int_{0}^{\epsilon} d \, dd \left| \psi(d)
\right|^{2} \hat{\sigma}(d,x^{\prime})}} + \stackrel{region 2}{\overbrace{\int_{\epsilon}^{\Delta} d \, dd \left| \psi(d)
\right|^{2} \hat{\sigma}(d,x^{\prime})}} + \stackrel{region 3}{\overbrace{\int_{\Delta}^{\infty} d \, dd \left| \psi(d)
\right|^{2} \hat{\sigma}(d,x^{\prime})}} \,.
\label{eq:proof2}
\end{equation}
Call the terms in Eq.~(\ref{eq:proof2}) regions 1,2, and 3 respectively.  For any given range of photon energies, one can 
choose sufficiently large $\Delta$ and sufficiently small $\epsilon$, that
regions 1 and 3 must give a negligible contribution to the over-all cross section.   We will justify this statement now.

First, in region 1, the cross section for the subprocess has the following behavior due to the first and last bulleted assumption above:
\begin{equation}
\hat{\sigma}_{1}(d,x^{\prime}) \sim d^{2} \left( E_{\gamma} \right) ^{\alpha},
\label{eq:proof3}
\end{equation}
where $\alpha$ is some positive real number.
Furthermore, the light cone wavefunction of the photon gets its energy dependence from the
leading behavior of the modified Bessel functions in Eq.~(\ref{eq:equation10}).  In the limit of $d << 1/m_{q}$, $\left| \psi(d) \right|^{2} \propto 1/d^{2}$.
Hence, in the limit defined by region 1, we have the following general energy dependence:
\begin{equation}
region \, 1 \sim \int_{0}^{\epsilon} d \, dd \frac{1}{d^{2}} d^{2} (E_{\gamma})^{\alpha} \sim \epsilon^{2} (E_{\gamma})^{\alpha + 1}.
\label{eq:proof4}
\end{equation}

Next we consider the other extreme: $d \rightarrow \infty$.  
Away from $d = 0$, Eq.~(\ref{eq:proof3}) shows that the cross section, $\hat{\sigma}(d,x^{\prime})$ for the subprocess 
rises very quickly to values that violate the unitarity constraint since we are considering UHE photons.  At a certain value of $d$, 
the growth of $\hat{\sigma}(d,x^{\prime})$ with $d$ must level out.  Within the dipole model, 
the growth of $\hat{\sigma}(d,x^{\prime})$ is flat with respect to variations in the transverse size of the
hadronic component in the limit that $d$ is large.  Call the upper limit of the basic cross section, $\hat{\sigma}_{max}$.  
Also, for $d \rightarrow \infty$, the Bessel functions give $\left| \psi(d) \right|^{2} \propto \frac{e^{-2 m_{q} d}}{d}$.  The energy dependence of the large size 
cross section can grow no faster than $\ln^{2} E_{\gamma}$ due to the Froissart bound (the second bulleted assumption above).
Thus, for region 3 we have,
\begin{equation}
region \, 3 \sim \int_{\Delta}^{\infty} d \, dd \frac{e^{-2 m_{q} d}}{d} \hat{\sigma}_{max} \ln^{2} (E_{\gamma}) \sim e^{-2 m_{q} \Delta} \ln^{2} E_{\gamma}.
\label{eq:proof5}
\end{equation}

For a particular
range of photon energies, we may always choose $\epsilon$ small enough, and $\Delta$ large enough that regions 1 and 2 give a negligible 
contribution to the total over-all integral in Eq.~(\ref{eq:proof2}).
From now on, assume that $\epsilon$ and $\Delta$ are always chosen, in each energy range, so that regions 1 and regions 2 are \emph{defined} to be negligibly small. 
Due to the general properties of the 
dipole model there is always a very small contribution from very large hadronic sizes ($d > \Delta$) that grows slowly with energy ($\sim \ln^{2} E_{\gamma}$), and 
there is always a small contribution from very small hadronic sizes whose contribution may grow very quickly (as a power of $E_{\gamma}$, according to the third
bulleted assumption) due to the fact that there will always be a 
contribution from extremely small sizes whose value of $\hat{\sigma}(d,x^{\prime})$ has not yet reached the unitarity limit at a given photon energy.  
We will now consider the rate of growth of the cross section that results from assuming that the cross section attains the maximum value allowed by unitarity for the largest
range of sizes possible within the general constraints of the dipole model.  Regions 1 and 3 give negligible contributions to the total integral, as discussed above, 
and the values of $\hat{\sigma}(d,x^{\prime})$ will be assumed to saturate the unitarity constraint for all values of $d$ outside of range of region 1.  This means that for region 2, the basic 
cross section (denoted by a subscript 2) has reached the maximum allowed value, $\hat{\sigma}_{max}$, in terms of its growth with $d$, and the rate of growth with $E_{\gamma}$ is the maximum allowed by the 
Froissart bound.  The cross section appearing in the integrand of region 2 then becomes,
\begin{equation}
\hat{\sigma}_{2}(d,x^{\prime}) \sim \hat{\sigma}_{max} \ln^{2} E_{\gamma}.
\label{eq:proof5bb}
\end{equation} 
However, the requirement that region 2 contains \emph{all} of the unitarity saturating contribution, and that region 1 contains a negligible contribution
 demands that we allow $\epsilon$ to have some 
energy dependence.  This is because, as Eq.~(\ref{eq:proof3}) shows, the basic cross section at small sizes may have as much as a quadratic 
$d$-dependence and potentially very rapid energy dependence.  
Therefore, at a small but fixed value of $d$, the basic cross section quickly rises from small values to unitarity violating
values with increasing energy.  However, region 1 by definition contains only the suppressed part of the integrand near $d = 0$, whereas
the unitarity saturating region should be associated entirely with region 2.  
As the energy of the photon is increased, therefore, we must continuously redefine  
region 1 so that the integrand of region 1 is confined to a smaller and smaller region around $d = 0$.   
This sort of behavior does not exist at $d \gtrsim \Delta$, because at such large values of hadronic size, the cross section only
increases as $\ln^{2} E_{\gamma}$ and there is almost no variation with hadronic size.  
Therefore, $\Delta$ is defined without any energy dependence ($\Delta$ may be given weak energy 
dependence, but that will only result in subleading powers of $\ln E_{\gamma}$ in the final result.)    
Equation~(\ref{eq:proof3}) tells us that the maximum rate at which $\epsilon$ may decrease at small $d$ is,
\begin{equation}
\epsilon(E_{\gamma}) \sim E_{\gamma}^{-(1 + \alpha)}.
\label{eq:proof5b}
\end{equation}
We thus write region 2 as,
\begin{equation}
region \, 2 \sim \int_{\epsilon(E_{\gamma})}^{\Delta} d \, dd \frac{1}{d^{2}} \hat{\sigma}_{max} \ln^{2} E_{\gamma} \sim \ln^{3} (E_{\gamma})
\label{eq:proof6}
\end{equation}
Here we have continued to use $1/d^{2}$ behavior for the squared photon wavefunction because this yields the fastest possible rate of divergence of the 
wavefunction at the lower end of the integral and thus yields the most conservative upper limit on the rate of growth.  Notice that after having given 
$\epsilon$ energy dependence, Eq.~(\ref{eq:proof4}) becomes,
\begin{equation}
region \, 1 \sim \int_{0}^{\epsilon(E_{\gamma})} d \, dd \frac{1}{d^{2}} d^{2} \ln^{2} (E_{\gamma}) \sim E_{\gamma}^{-(1 + \alpha)}.
\label{eq:proof7}
\end{equation}
Thus we have established the behavior of each of the regions in Eq.~(\ref{eq:proof1}).  Regions 1 yields a vanishing contribution
to the total integral as $E_{\gamma} \rightarrow \infty$ and region 3 has energy dependence $\lesssim \ln^{2} E_{\gamma}$ whereas
region 2 has energy dependence, $\lesssim \ln^{3}(E_{\gamma})$. 
Taking the leading behavior in Eq.~(\ref{eq:proof1}) therefore gives,
\begin{equation}
\sigma(\gamma \,\, hadron \rightarrow hadrons) \lesssim  Constant \times \ln^{3} E_{\gamma}.
\label{eq:proof8}
\end{equation} 
Equation~(\ref{eq:proof8}) applies for each active flavor individually and thus for the sum of flavors.  A possible way that
the rate of growth at ultra-high energies violates Eq.~(\ref{eq:proof8}) in spite of the unitarity limit being saturated for each flavor would be
for there to be a large proliferation of new active flavors at ultra high energies.


\begin{thebibliography}{99}
\bibitem{footnote3} In principle one can write a more accurate formula which would take into account
deviations of $\frac{ \langle \hat{\sigma}^{n} \rangle}{\sigma_{\gamma N}}$ from 
$\sigma_{eff}^{n-1}$ for $n\ge 3$. However numerically these effects are small especially for the light nuclei where double scattering gives a dominant contribution.
\end{thebibliography}
\end{document}